# Chapter 1

# High Luminosity Large Hadron Collider HL-LHC

*G. Apollinari[1], O. Brüning[2], T. Nakamoto[3] and L. Rossi[2]\**

[1]Fermi National Accelerator Laboratory, Batavia, USA
[2]CERN, Accelerator & Technology Sector, Geneva, Switzerland
[3]KEK, Tsukuba, Japan

## 1 High Luminosity Large Hadron Collider HL-LHC

### 1.1 Introduction

The Large Hadron Collider (LHC) was successfully commissioned in 2010 for proton–proton collisions with a 7 TeV centre-of-mass energy and delivered 8 TeV centre-of-mass proton collisions from April 2012 to the end of 2013. The LHC is pushing the limits of human knowledge, enabling physicists to go beyond the Standard Model. The announcement given by CERN on 4 July 2012 about the discovery of a new boson at about 125 GeV, the long-awaited Higgs particle, is the first fundamental discovery, hopefully the first of a series that LHC can deliver.

It is a remarkable era for cosmology, astrophysics and high energy physics and the LHC is at the forefront of attempts to understand the fundamental nature of the universe. The discovery of the Higgs boson in 2012 is undoubtedly a major milestone in the history of physics. Beyond this, the LHC has the potential to go on and help answer some of the key questions of the age: the existence, or not, of supersymmetry; the nature of dark matter; the existence of extra dimensions. It is also important to continue to study the properties of the Higgs – here the LHC is well placed to do this in exquisite detail.

Thanks to the LHC, Europe has decisively regained world leadership in High Energy Physics (HEP), a key sector of knowledge and technology. The LHC can continue to act as catalyst for a global effort unrivalled by any other branch of science: out of the 10000 CERN users, more than 7000 are scientists and engineers using the LHC, half of which are from countries outside the EU.

The LHC will remain the most powerful accelerator in the world for at least the next two decades. Its full exploitation is the highest priority of the European Strategy for particle physics. This strategy has been adopted by the CERN Council, and is a reference point for the Particle Physics Strategy of the US and, to a certain extent, Japan. To extend its discovery potential, the LHC will need a major upgrade in the 2020s to increase its luminosity (and thus collision rate) by a factor of five beyond its design value. The integrated luminosity design goal is an increase by a factor of ten. As a highly complex and optimized machine, such an upgrade must be carefully studied. The necessary developments will require about 10 years to prototype, test and realize. The novel machine configuration, the High Luminosity LHC (HL-LHC), will rely on a number of key innovative technologies representing exceptional technological challenges. These include among others: cutting-edge 11–12 T superconducting magnets; very compact with ultra-precise phase control superconducting cavities for beam rotation; new technology for beam collimation; and long high-power superconducting links with zero energy dissipation.

---

\* Corresponding author: Lucio.Rossi@cern.ch

HL-LHC federates the efforts and R&D of a large international community towards the ambitious HL-LHC objectives and contributes to establishing the European Research Area (ERA) as a focal point of global research cooperation and a leader in frontier knowledge and technologies. HL-LHC relies on strong participation from various partners, in particular from leading US and Japanese laboratories. This participation will be required for the execution of the construction phase as a global project. In particular, the US LHC Accelerator R&D Program (LARP) has developed some of the key technologies for the HL-LHC, such as the large-aperture niobium–tin ($Nb_3Sn$) quadrupoles and the crab cavities. The proposed governance model is tailored accordingly and should pave the way for the organization of the construction phase.

## 1.2    HL-LHC in a nutshell

The LHC baseline programme until 2025 is shown schematically in Figure 1-1. After entering into the nominal energy regime of 13–14 TeV centre-of-mass energy in 2015, it is expected that the LHC will reach the design luminosity of $1 \times 10^{34}$ cm$^{-2}$ s$^{-1}$. This peak value should give a total integrated luminosity of about 40 fb$^{-1}$ per year. In the period 2015–2022 the LHC will hopefully further increase the peak luminosity. Margins in the design of the nominal LHC are expected to allow, in principle, about two times the nominal design performance. The baseline programme for the next ten years is depicted in Figure 1-1, while Figure 1–2 shows the possible evolution of peak and integrated luminosity.

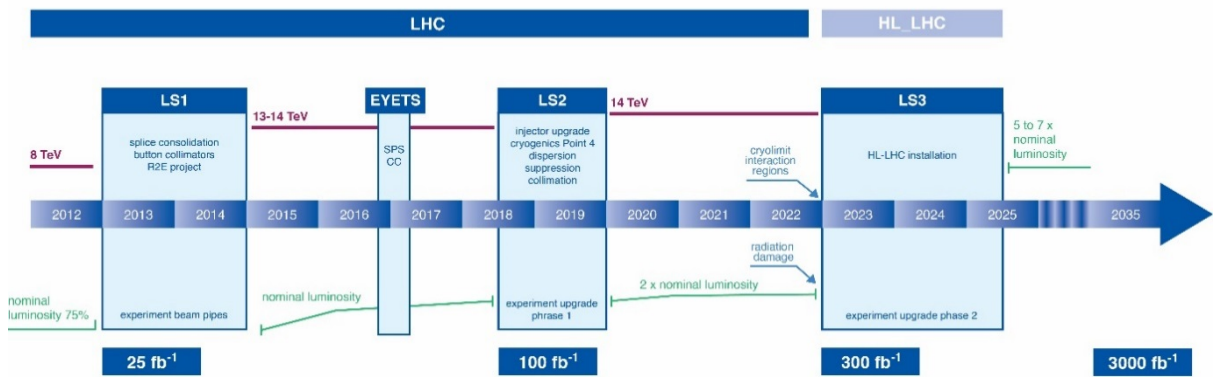

Figure 1-1: LHC baseline plan for the next decade and beyond showing the energy of the collisions (upper red line) and luminosity (lower green lines). The first long shutdown (LS1) in 2013–2014 will allow the design parameters of beam energy and luminosity to be reached. The second long shutdown (LS2) in 2018–2019, will consolidate luminosity and reliability as well as the upgrading of the LHC injectors. After LS3, 2023–2025, the machine will be in the High Luminosity configuration (HL-LHC).

After 2020 the statistical gain in running the accelerator without a significant luminosity increase beyond its design value will become marginal. The running time necessary after 2020 to halve the statistical error in measurements will be more than ten years. Therefore, to maintain scientific progress and to explore its full capacity, the LHC will need to have a decisive increase of its luminosity. This is why, when the CERN Council adopted the European Strategy for particle physics in 2006 [1], its first priority was agreed to be 'to fully exploit the physics potential of the LHC. A subsequent major luminosity upgrade, motivated by physics results and operation experience, will be enabled by focused R&D'. The European Strategy for particle physics has been integrated into the European Strategy Forum on Research Infrastructures (ESFRI) Roadmap of 2006, as has the update of 2008 [2]. The priority to fully exploit the potential of the LHC has recently been confirmed as the *first priority* among the 'High priority large-scale scientific activities' in the new European Strategy for particle physics – Update 2013 [3]. This update was approved in Brussels on 30 May 2013 with the following wording: 'Europe's top priority should be the exploitation of the full potential of the LHC, including the high luminosity upgrade of the machine and detectors with a view to collecting ten times more data than in the initial design, by around 2030'.

The importance of the LHC luminosity upgrade for the future of high energy physics has been also recently re-affirmed by the May 2014 recommendation by the Particle Physics Project Prioritization Panel (P5) to the High Energy Physics Advisory Panel (HEPAP), which in turn advises the US Department of Energy (DOE) [4]. The recommendation, a critical step in the updating of the US strategy for HEP, states the following: 'Recommendation 10: The LHC upgrades constitute our highest-priority near-term large project'.

In Japan, the 2012 report of a subcommittee in the HEP community concluded that an $e^+e^-$ linear collider and a large-scale neutrino detector would be the core projects in Japan, with the assumption that the LHC and its upgrade are pursued de facto. The updated KEK roadmap in 2013 states that 'The main agenda at LHC/ATLAS is to continually participate in the experiment and to take a proactive initiative in upgrade programmes within the international collaboration at both the accelerator and detector facilities.' Following these supports, The ATLAS-Japan group has undertaken intensive R&D on the detector upgrades and the KEK cryogenic group has started the R&D upon the LHC separation dipole magnet.

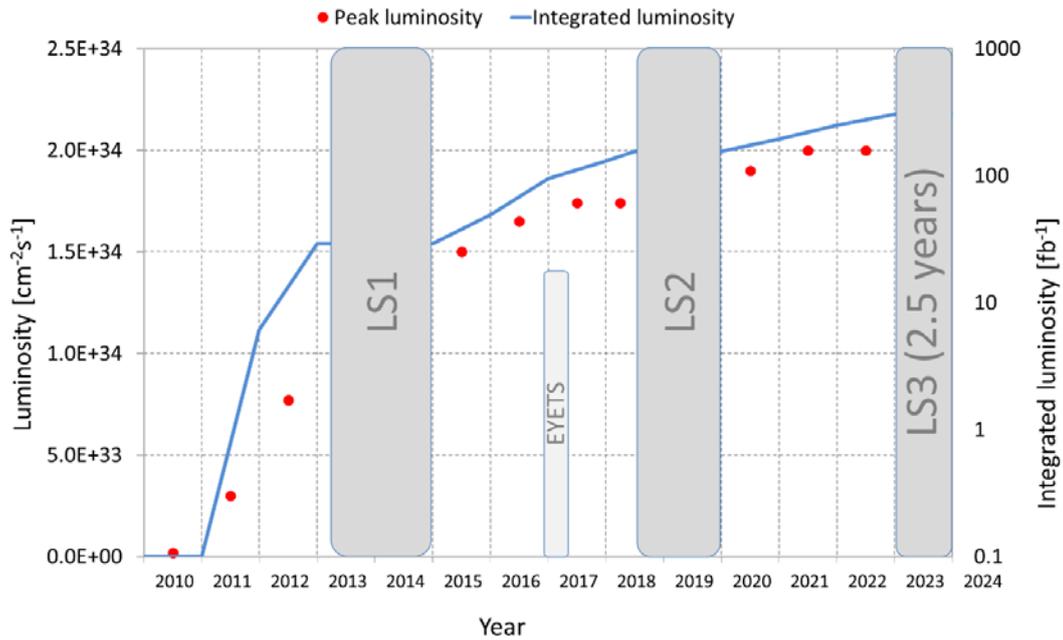

Figure 1-2: LHC luminosity plan for the next decade, both peak (red dots) and integrated (blue line). Main shutdown periods are indicated.

In this context, at the end of 2010 CERN created the High Luminosity LHC (HL-LHC) project [5]. Started as a design study, and after the approval of the CERN Council of 30 May 2013 and the insertion of the budget in the CERN Medium Term Plan approved by Council in June 2014, the HL-LHC has become CERN's major construction project for the next decade.

The main objective of the High Luminosity LHC design study is to determine a set of beam parameters and the hardware configuration that will enable the LHC to reach the following targets:

- A peak luminosity of $5 \times 10^{34}$ cm$^{-2}$ s$^{-1}$ with levelling, allowing:

- An integrated luminosity of 250 fb$^{-1}$ per year with the goal of 3000 fb$^{-1}$ in about a dozen years after the upgrade. This integrated luminosity is about ten times the expected luminosity reach of the first twelve years of the LHC lifetime.

The overarching goals are the installation of the main hardware for the HL-LHC and the commissioning of the new machine configuration during LS3, scheduled for 2023–2025, while taking all actions to assure a high efficiency in operation until 2035.

Actually, during the last year, the necessity emerged of aiming at an enhanced goal in terms of annual integrated luminosity. If the target of 3000 fb$^{-1}$ should be reached by around 2035, as inferred by the European Strategy Update, the nominal goal of 250 fb$^{-1}$/year as fixed above is probably not adequate. However, since all equipment is being designed with a margin of 50%, regarding reaching the required luminosity, we are defining the concept of *ultimate parameters*. By using these margins we should be able to push our machine to about 7–7.5 × 10$^{34}$ cm$^{-2}$ s$^{-1}$ of peak, levelling luminosity, therefore of course increasing the total pile-up in the detectors up to 200. This luminosity level should enable the collect of up to 300–350 fb$^{-1}$/year. Also, in terms of total integrated luminosity, we think we can define an ultimate value of about 4000 fb$^{-1}$. It must be said that while at first examination there is no showstopper for these performances, the ultimate parameters are not yet consolidated as the nominal parameters. Therefore, they will be thoroughly scrutinized and consolidated for the next version of the technical design report.

All of the hadron colliders in the world before the LHC have produced a combined total integrated luminosity of about 10 fb$^{-1}$. The LHC delivered nearly 30 fb$^{-1}$ by the end of 2012 and should reach 300 fb$^{-1}$ in its first 13–15 years of operation. The High Luminosity LHC is a major, extremely challenging, upgrade. For its successful realization, a number of key novel technologies have to be developed, validated, and integrated. The work was initiated quite early: ideas were circulating at the beginning of LHC construction [6] and this continued throughout construction [7]. From 2003, LARP (see Section 1.3.2) has been the main and continuous motor for technological development devoted to the LHC upgrade. After a period during which the upgrade was conceived in two phases, all studies were unified in 2010 under the newly formed High Luminosity Project. The first step consisted in launching a Design Study under the auspices of EC-FP7 with the nickname 'HiLumi LHC', which, following approval by the EC in 2011, has been instrumental in initiating a new global collaboration for the LHC matching the spirit of the worldwide user community of the LHC experiments.The High Luminosity LHC project is working in close collaboration with the CERN project for the LHC Injector complex Upgrade (LIU) [8], the companion ATLAS and CMS upgrade projects of 2018–2019 and 2023–2025 and the upgrade foreseen in 2018–2019 for both LHCb and Alice.

### 1.2.1 Luminosity

The (instantaneous) luminosity *L* can be expressed as:

$$L = \gamma \frac{n_b N^2 f_{rev}}{4\pi \beta^* \varepsilon_n} R; \quad R = 1/\sqrt{1 + \frac{\theta_c \sigma_z}{2\sigma}} \tag{1-1}$$

where $\gamma$ is the proton beam energy in unit of rest mass; $n_b$ is the number of bunches per beam: 2808 (nominal LHC value) for 25 ns bunch spacing; *N* is the bunch population. $N_{nominal\ 25\ ns}$: 1.15×10$^{11}$ p ($\Rightarrow$0.58 A of beam current at 2808 bunches); $f_{rev}$ is the revolution frequency (11.2 kHz); $\beta^*$ is the beam beta function (focal length) at the collision point (nominal design 0.55 m); $\varepsilon_n$ is the transverse normalized emittance (nominal design: 3.75 μm); *R* is a luminosity geometrical reduction factor (0.85 at a $\beta^*$ *of* 0.55 m of, down to 0.5 at 0.25 m); $\theta_c$ is the full crossing angle between colliding beam (285 μrad as nominal design); and $\sigma$, $\sigma_z$ are the transverse and longitudinal r.m.s. sizes, respectively (nominally 16.7 μm and 7.55 cm, respectively)

With the nominal parameter values shown above, a luminosity of 1 × 10$^{34}$ cm$^{-2}$ s$^{-1}$ is obtained, with an average pile-up (number of events in the same bunch crossing) of $\mu$ = 27 (although $\mu$ = 19 was the original forecast at LHC approval due to uncertainties in the total proton cross-section at higher energies).

### 1.2.2 Present luminosity limitations and hardware constraints

There are various expected limitations to an increase in luminosity, either from beam characteristics (injector chain, beam impedance and beam–beam interactions in the LHC) or from technical systems. Mitigation of potential performance limitations arising from the LHC injector complex are addressed by the LIU project mentioned above, which should be completed in 2019 (after LS2). Any potential limitations coming from the LHC injector complex aside, it is expected that the present LHC will reach a performance limitation from the beam current, from cleaning efficiency with 350 MJ beam stored energy, from e-clouds effects, from the maximum available cooling in the triplet magnets, from the magnet aperture ($\beta^*$ limit) and from the acceptable pile-up level. The ultimate value of bunch population with the nominal LHC should enable a peak luminosity of around $2 \times 10^{34}$ cm$^{-2}$ s$^{-1}$ to be reached. Any further performance increase of the LHC will require significant hardware and beam parameter modifications with respect to the design LHC configuration.

Before discussing the new configuration it is useful to recall the systems that need to be changed, and possibly improved, because they become vulnerable to breakdown and accelerated aging, or because they may become a bottleneck for operation in a higher radiation environment. This goes well beyond the ongoing basic consolidation.

- Inner triplet magnets. After about 300 fb$^{-1}$ some components of the inner triplet quadrupoles and their corrector magnets will have received a dose of 30 MGy, entering into the region of possible radiation damage. The quadrupoles may withstand a maximum of 400 fb$^{-1}$ to 700 fb$^{-1}$, but some corrector magnets of nested type are likely to have already failed at 300 fb$^{-1}$. Actual damage must be anticipated because the most likely failure mode is through sudden electric breakdown, entailing serious and long repairs. Thus the replacement of the triplet magnets must be envisaged before damage occurs. Replacement of the low-beta triplet is a long intervention, requiring a one- to two-year shutdown and must be coupled with major detector upgrades.

- Cryogenics. To increase intervention flexibility and machine availability it is planned to install a new cryogenics plant for a full separation between superconducting RF (SCRF) and magnet cooling. In the long term, the cooling of the inner triplets and matching section magnets must be separated from the arc magnets. This would avoid the need to warm-up an entire arc in the case of triplet region intervention.

- Collimation. The collimation system has been designed for the first operation phase of the LHC. The present system was optimized for robustness and will need an upgrade that takes into account the need for the lower impedance required for the planned increased beam intensities. A new configuration will also be required to protect the new triplets in IR1 and IR5.

- Also requiring special attention are the dispersion suppressor (DS) regions, where a leakage of off-momentum particles into the first and second main superconducting dipoles has been already identified as a possible LHC performance limitation. The most promising concept is to substitute an LHC main dipole with dipoles of equal bending strength (~120 T·m) obtained by a higher field (11 T) and shorter length (11 m) than those of the LHC dipoles (8.3 T and 14.2 m). The room gained is sufficient for the installation of special collimators.

- Radiation to electronics (R2E) and superconducting links for the remote powering of cold circuits. Considerable effort is being made to study how to replace the radiation-sensitive electronics boards of the power converter system with radiation-hard cards. A complementary solution is also being pursued for special zones. This would entail removal of the power converters and associated electrical feedboxes (DFBs), delicate equipment presently in line with the continuous cryostat) out of the tunnel, possibly to the surface. LHC availability should be improved. In particular in LHC P7, where a set of 600 A power converters are placed near the betatron cleaning collimators, removal will be to a lateral tunnel because the surface is not accessible. Displacement of power converters to distant locations is possible only thanks to a novel technology: superconducting links (SCLs) made from YBCO or Bi-2223 High Temperature Superconductors (HTS) or $MgB_2$ superconductors.

- Quench Protection System (QPS), machine protection and remote manipulation. Other systems will potentially become problematic, along with the aging of the machine and the radiation level that comes with higher performance levels of 40 fb$^{-1}$ to 60 fb$^{-1}$ per year:

  o QPS for the superconducting magnets, based on a design that is almost 20 years old.

  o Machine protection: improved robustness to mis-injected beams, kicker sparks and asynchronous dumps will be required. The kicker system is, with collimation and the injection beam stopper, the main shield against severe beam-induced damage. The kicker systems, along with the system will need renovation after 2020.

- o Remote manipulation: the level of activation from 2020 onwards, and perhaps even earlier, requires careful study and the development of special equipment to allow the replacement of collimators, magnets, vacuum components, etc., according to the 'as low as reasonably achievable' (ALARA) principle. While full robotics is difficult to implement, given the conditions on the ground, remote manipulation, enhanced reality and supervision are the key to minimizing the radiation doses sustained during interventions.

### 1.2.3 Luminosity levelling, availability

Both the consideration of energy deposition by collision debris in the interaction region magnets, and the necessity to limit the peak pile-up in the experimental detector, impose an a priori limitation upon peak luminosity. The consequence is that HL-LHC operation will have to rely on luminosity levelling. As shown in Figure 1-3(a), the luminosity profile without levelling quickly decreases from the initial peak value due to 'luminosity burn' (protons consumed in the collisions). The collider is designed to operate with a constant luminosity at a value below the virtual maximum luminosity. The average luminosity achieved is almost the same as that without levelling, see Figure 1-3(b). The advantage, however, is that the maximum peak luminosity is lower.

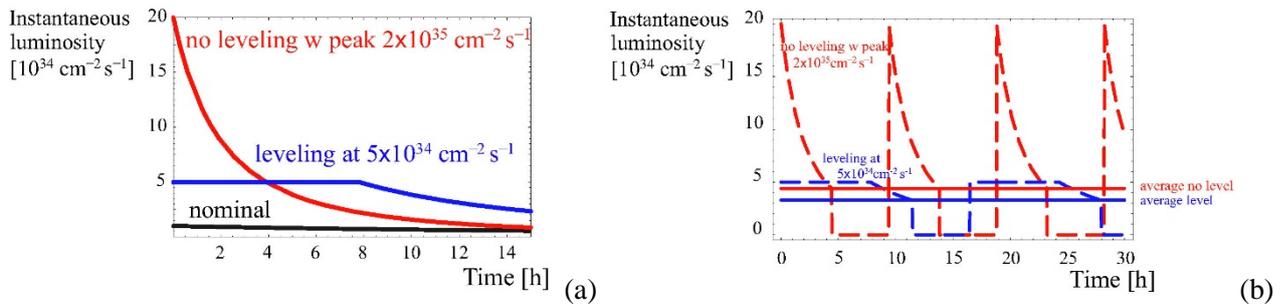

Figure 1-3: (a) Luminosity profile for a single long fill: starting at nominal peak luminosity (black line), with upgrade no levelling (red line), with levelling (blue line). (b) Luminosity profile with optimized run time, without and with levelling (blue and red dashed lines), and average luminosity in both cases (solid lines).

Because of the levelled luminosity limit, to maximize the integrated luminosity one needs to maximize the fill length. This can be achieved by maximizing the injected beam current. Other key factors for maximizing the integrated luminosity and obtaining the required 3 fb$^{-1}$/day (see Figure 1-4) are a short average machine turnaround time, an average operational fill length that exceeds the luminosity levelling time, and good overall machine efficiency. The machine efficiency is essentially the available time for physics after downtime for fault recovery is taken into account. Closely related is the physics efficiency – the fraction of time per year spent actually providing collisions to the experiments. For integrated luminosity the efficiency counts almost as much as the virtual peak performance.

The HL-LHC with 160 days of physics operation a year needs a physics efficiency of about 40%. The overall LHC efficiency during the 2012 run, without luminosity levelling, was around 37%. The requirement of an efficiency higher than the one of the present LHC, with a (levelled) luminosity five times that of nominal, will be a real challenge. The project must foresee a vigorous consolidation for the high intensity and high luminosity regime: the High Luminosity LHC must also be a high availability LHC.

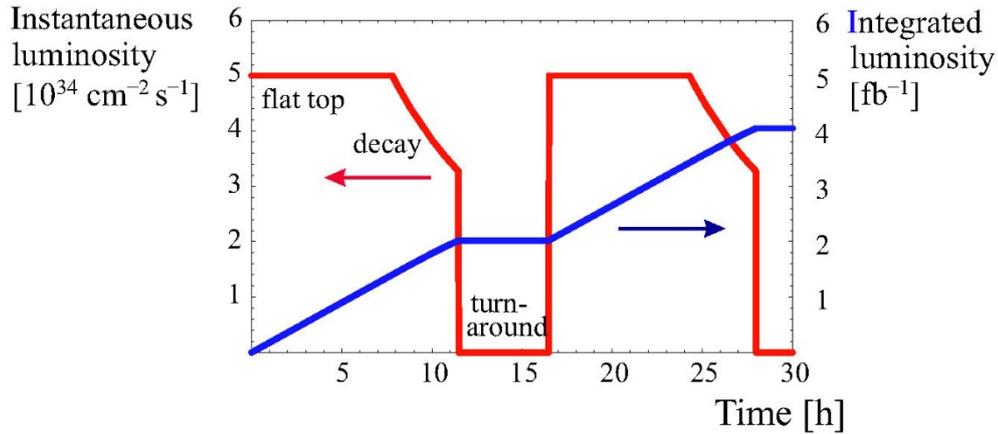

Figure 1-4: Luminosity cycle for HL-LHC with levelling and a short decay (optimized for integrated luminosity).

1.2.4 HL-LHC parameters and main systems for the upgrade

Table 1-1 lists the main parameters foreseen for high luminosity operation. The 25 ns bunch spacing is the baseline operation mode; however, 50 ns bunch spacing is kept as a possible alternative in case the e-cloud or other unforeseen effects undermine 25 ns performance. A slightly different parameter set at 25 ns (batch compression and beam merging scheme (BCMS)) with very small transverse beam emittance is also shown and might be interesting for HL-LHC operation in case operation with high beam intensities results in unforeseen emittance blow-up.

Table 1-1: High Luminosity LHC parameters

| Parameter | Nominal LHC (design report) | HL-LHC 25 ns (standard) | HL-LHC 25 ns (BCMS) | HL-LHC 50 ns |
|---|---|---|---|---|
| Beam energy in collision [TeV] | 7 | 7 | 7 | 7 |
| $N_b$ | $1.5 \times 10^{11}$ | $2.2 \times 10^{11}$ | $2.2 \times 10^{11}$ | $3.5 \times 10^{11}$ |
| $n_b$ | 2808 | 2748 | 2604 | 1404 |
| Number of collisions in IP1 and IP5 | 2808 | 2736* | 2592 | 1404 |
| $N_{tot}$ | $3.2 \times 10^{14}$ | $6 \times 10^{14}$ | $5.7 \times 10^{14}$ | $4.9 \times 10^{14}$ |
| Beam current [A] | 0.58 | 1.09 | 1.03 | 0.89 |
| Crossing angle [μrad] | 285 | 590 | 590 | 590 |
| Beam separation [$\sigma$] | 9.4 | 12.5 | 12.5 | 11.4 |
| $\beta^*$ [m] | 0.55 | 0.15 | 0.15 | 0.15 |
| $\varepsilon_n$ [μm] | 3.75 | 2.50 | 2.50 | 3 |
| $\varepsilon_L$ [eVs] | 2.50 | 2.50 | 2.50 | 2.50 |
| r.m.s. energy spread | $1.13 \times 10^{-4}$ | $1.13 \times 10^{-4}$ | $1.13 \times 10^{-4}$ | $1.13 \times 10^{-4}$ |
| r.m.s. bunch length | $7.55 \times 10^{-2}$ | $7.55 \times 10^{-2}$ | $7.55 \times 10^{-2}$ | $7.55 \times 10^{-2}$ |
| IBS horizontal [h] | 80–106 | 18.5 | 18.5 | 17.2 |
| IBS longitudinal [h] | 61–60 | 20.4 | 20.4 | 16.1 |
| Piwinski parameter | 0.65 | 3.14 | 3.14 | 2.87 |
| Geometric loss factor $R_0$ without crab cavity | 0.836 | 0.305 | 0.305 | 0.331 |
| Geometric loss factor $R_1$ with crab cavity | (0.981) | 0.829 | 0.829 | 0.838 |
| Beam–beam/IP without crab cavity | $3.1 \times 10^{-3}$ | $3.3 \times 10^{-3}$ | $3.3 \times 10^{-3}$ | $4.7 \times 10^{-3}$ |
| Beam–beam/IP with crab cavity | $3.8 \times 10^{-3}$ | $1.1 \times 10^{-2}$ | $1.1 \times 10^{-2}$ | $1.4 \times 10^{-2}$ |
| Peak luminosity without crab cavity [$cm^{-2} s^{-2}$] | $1.00 \times 10^{34}$ | $7.18 \times 10^{34}$ | $6.80 \times 10^{34}$ | $8.44 \times 10^{34}$ |
| Virtual luminosity with crab cavity, $L_{peak} \times R_1/R_0$ [$cm^{-2} s^{-2}$] | $(1.18 \times 10^{34})$ | $19.54 \times 10^{34}$ | $18.52 \times 10^{34}$ | $21.38 \times 10^{34}$ |
| Events/crossing without levelling and without crab cavity | 27 | 198 | 198 | 454 |
| Levelled luminosity [cm$^{-2}$ s$^{-2}$] | - | $5.00 \times 10^{34\dagger}$ | $5.00 \times 10^{34}$ | $2.50 \times 10^{34}$ |
| Events/crossing (with levelling and without crab cavities for HL-LHC) | 27 | 138 | 146 | 135 |
| Peak line density of pile-up event [event/mm] (maximum over stable beams) | 0.21 | 1.25 | 1.31 | 1.20 |
| Levelling time [h] (assuming no emittance growth) | - | 8.3 | 7.6 | 18.0 |
| Number of collisions in IP2/IP8 | 2808 | 2452/2524$^\ddagger$ | 2288/2396 | 0$^{**}$/1404 |
| $N_b$ at SPS extraction$^{\dagger\dagger}$ | $1.20 \times 10^{11}$ | $2.30 \times 10^{11}$ | $2.30 \times 10^{11}$ | $3.68 \times 10^{11}$ |
| $n_b$/injection | 288 | 288 | 288 | 144 |
| $N_{tot}$/injection | $3.46 \times 10^{13}$ | $6.62 \times 10^{13}$ | $6.62 \times 10^{13}$ | $5.30 \times 10^{13}$ |
| $\varepsilon_n$ at SPS extraction [μm]$^\ddagger$ | 3.40 | 2.00 | <2.00$^{***}$ | 2.30 |

*Assuming one less batch from the PS for machine protection (pilot injection, Transfer line steering with 12 nominal bunches) and non-colliding bunches for experiments (background studies, etc.). Note that due to RF beam loading the abort gap length must not exceed the 3 μs design value.

$^\dagger$For the design of the HL-LHC systems (collimators, triplet magnets, etc.), a margin of 50% on the stated peak luminosity (corresponding to the ultimate levelled luminosity) has been agreed.

$^\ddagger$The total number of events/crossing is calculated with an inelastic cross-section of 85 mb (also for nominal), while 100 mb is still assumed for calculating the proton burn off and the resulting levelling time.

$^{**}$The lower number of collisions in IR2/8 compared to the general-purpose detectors is a result of the agreed filling scheme, aiming as much as possible at a democratic sharing of collisions between the experiments.

$^{\dagger\dagger}$An intensity loss of 5% distributed along the cycle is assumed from SPS extraction to collisions in the LHC.

$^\ddagger$A transverse emittance blow-up of 10–15% on the average H/V emittance in addition to that expected from intra-beam scattering (IBS) is assumed (to reach 2.5 μm of emittance in collision for 25 ns operation).

$^{***}$For the BCMS scheme emittances down to 1.7 μm have already been achieved at LHC injection, which might be used to mitigate excessive emittance blow-up in the LHC during injection and ramp An upgrade should provide the potential for performance over a wide range of parameters, and eventually the machine and experiments will find the best practical set of parameters in actual operations.

Beam current and brightness: the total beam current may be a hard limit in the LHC since many systems are affected by this parameter: RF power system and RF cavities, collimation, cryogenics, kickers, vacuum, beam diagnostics, QPS, etc. Radiation effects aside, all systems have been designed in principle for $I_{beam}$ = 0.86 A, the so-called 'ultimate' beam current. However the ability to go to the ultimate limit is still to be experimentally demonstrated and the HL-LHC will need to go 30% beyond ultimate with 25 ns bunch spacing.

For the HL-LHC there is a need to increase the beam brightness, a beam characteristic that must be maximized at the beginning of beam generation and then preserved throughout the entire injector chain and in LHC itself. The LIU project has as its primary objective increasing the number of protons per bunch by a factor of two above the nominal design value while keeping emittance at the present low value.

$\beta^*$ and cancelling the reduction factor R. A classical route for a luminosity upgrade is to reduce $\beta^*$ by means of stronger and larger aperture low-$\beta$ triplet quadrupoles. However a reduction in $\beta^*$ values implies not only larger beam sizes in the triplet magnets but also an increase in crossing angle. The increased crossing angle in turn requires even larger aperture triplet magnets, a larger aperture D1 (first separation dipole) and further modifications to the matching section. It also reduces the luminous region size and thus the gain in peak luminosity.

Stronger chromatic aberrations coming from the larger $\beta$-functions inside the triplet magnets may furthermore exceed the strength of the existing correction circuits. The peak $\beta$-function is also limited by the possibility to match the optics to the regular beta functions of the arcs. A previous study has shown that in the nominal LHC the practical limit for $\beta^*$ is 30 cm to 40 cm cf. the nominal 55 cm. However, a novel scheme called Achromatic Telescopic Squeeze (ATS) uses the adjacent arcs as enhanced matching sections. The increase of the beta-functions in these arcs can boost, at constant strength, the efficiency of the arc correction circuits. In this way a $\beta^*$ value of 15 cm can be envisaged, and flat optics with a $\beta^*$ as low as 5 cm in the plane perpendicular to the crossing plane could be realized. For such a $\beta^*$ reduction the triplet quadrupoles need to double their aperture, and require a peak field 50% above the present LHC. This implies the use of new, advanced, superconducting technology based on Nb$_3$Sn.

The drawback of very small $\beta^*$ is that it requires a larger crossing angle. This causes a reduction of the geometrical luminosity reduction factor R. In Figure 1-5 the reduction factor is plotted vs. $\beta^*$ values.

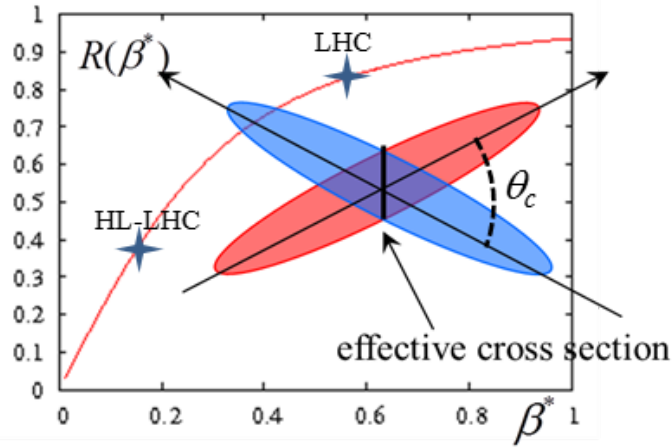

Figure 1-5: Behaviour of geometrical luminosity reduction factor vs. $\beta^*$ for a constant normalized beam separation with the indication of two operational points: nominal LHC and HL-LHC. The bunch crossing sketch shows the reduction mechanism.

Various methods can be employed to at least partially counteract this effect. The most efficient and elegant solution for compensating the geometric reduction factor is the use of special superconducting RF crab cavities, capable of generating transverse electric fields that rotate each bunch longitudinally by $\theta_c/2$, such that they effectively collide head on, overlapping perfectly at the collision points, as illustrated in Figure 1-6. Crab cavities allow access to the full performance of the small $\beta^*$ values offered by the ATS scheme and the larger triplet quadrupole magnets. While the primary function of the crab cavities is to boost the virtual peak luminosity, they can also be used in combination with dynamic $\beta^*$ variation during the fill. This would allow optimization of the size of the luminous region and thus the pile-up density through the fill. Finally, the crab cavities can be used to tilt the bunches in a direction perpendicular to the plane of crossing, providing pile-up control and an additional handle for luminosity levelling through the so-called 'crab-kissing' scheme.

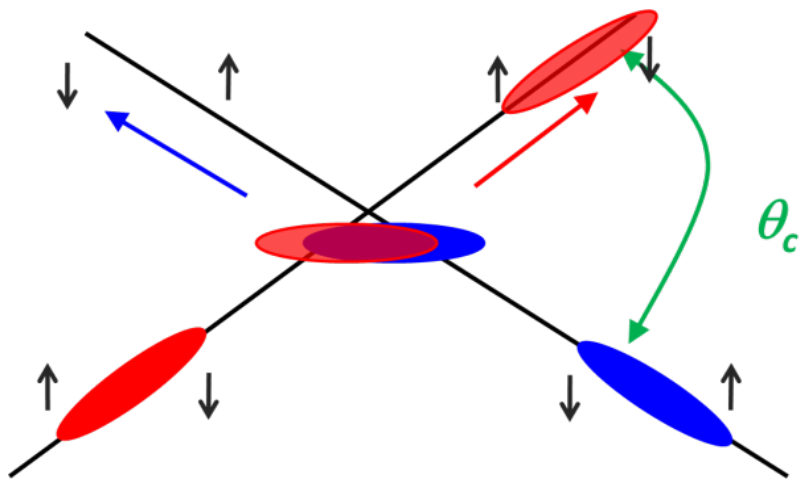

Figure 1-6: Effect of the crab cavity on the beam (small arrows indicate the torque on the bunch generated by the transverse RF field).

The layout and main hardware modifications required to produce the parameters listed in Table 1-1 are described in Chapter 2 of this report.

Given the yearly and long-term operations schedule, the targets of 250 fb$^{-1}$ per year and 3000 fb$^{-1}$ by the mid-2030s are very challenging. If the performance of the HL-LHC can go beyond the design levelled luminosity value of $L_{peak} = 5 \times 10^{34}$ cm$^{-2}$ s$^{-1}$ then these targets become more reasonable. Indeed, all systems will be designed with some margin. If the behaviour of the machine is such as to allow the utilization of these margins, and if the upgraded detectors will accept a higher pile-up, up to 200, then the performance could eventually reach $7.5 \times 10^{34}$ cm$^{-2}$ s$^{-1}$ with levelling. With a performance of 300 fb$^{-1}$/year, this would allow almost 4000 fb$^{-1}$ to be obtained by 2037, as shown in Figure 1-7.

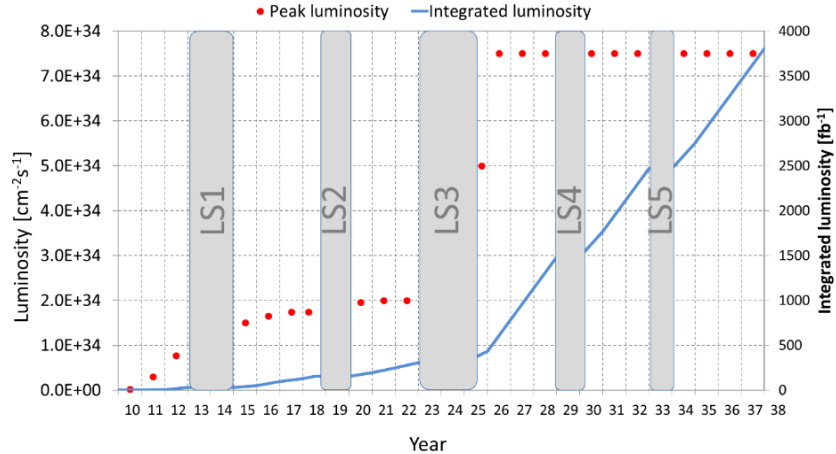

Figure 1-7: Forecast for peak luminosity (red dots) and integrated luminosity (blue line) in the HL-LHC era, for the case of ultimate HL-LHC parameters. Note that for the sake of simplicity there is no learning curve for luminosity after LS3.

1.2.5 Planning and costings

The HL-LHC schedule aims at the installation of the main HL-LHC hardware during LS3, together with the final upgrade of the experimental detectors (the so-called Phase II upgrade). However, a few items like the new cryogenic plant for P4, the 11 T dipole for DS collimation in P2 (for ions) and the SC links in P7 would be installed during LS2.

The HL-LHC schedule is based on the following milestones:

- 2014: Preliminary Design Report (PDR);
- 2015: End of design phase, release of the first Technical Design Report (TDR);
- 2016: Proof of main hardware components on test benches;
- 2017: Testing of prototypes (including crab cavity test in SPS) and release of TDR v1;
- 2017–2021: Construction and test of long-lead hardware components (e.g. magnets, crab cavities, SC links, collimators);
- 2018–2019: LS2 – Installation of cryo-plant P4, DS collimators (11 T) in P2, SC link in P7;
- 2020–2022: String test of inner triplet;
- 2023–2025: LS3 – Main installation (new magnets, crab cavities, cryo-plants, collimators, absorbers, etc.) and commissioning.

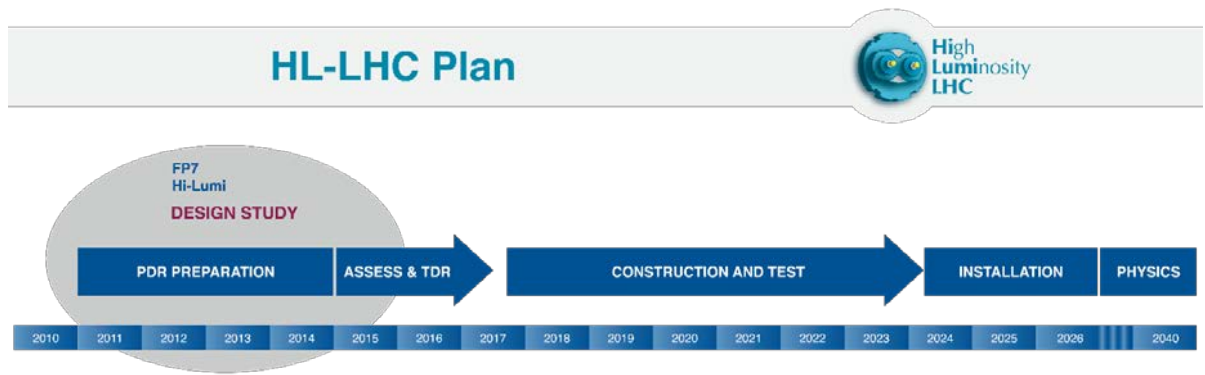

Figure 1-8: Schematic representation of the main HL-LHC milestones

The preliminary cost-to-completion (CtC) of the full HL-LHC project amounts to about 830 MCHF for materials (CERN accounting). A coarse evaluation of personnel requirements amounts to more than 1000 fulltime equivalent (FTE) years. The cost-to-completion does not include the civil engineering works for the underground infrastructures (presently under evaluation) and non-baseline systems such as the long-range beam–beam compensators, the RF harmonic system, and the related infrastructure. The budget profile is shown in Figure 1-9.

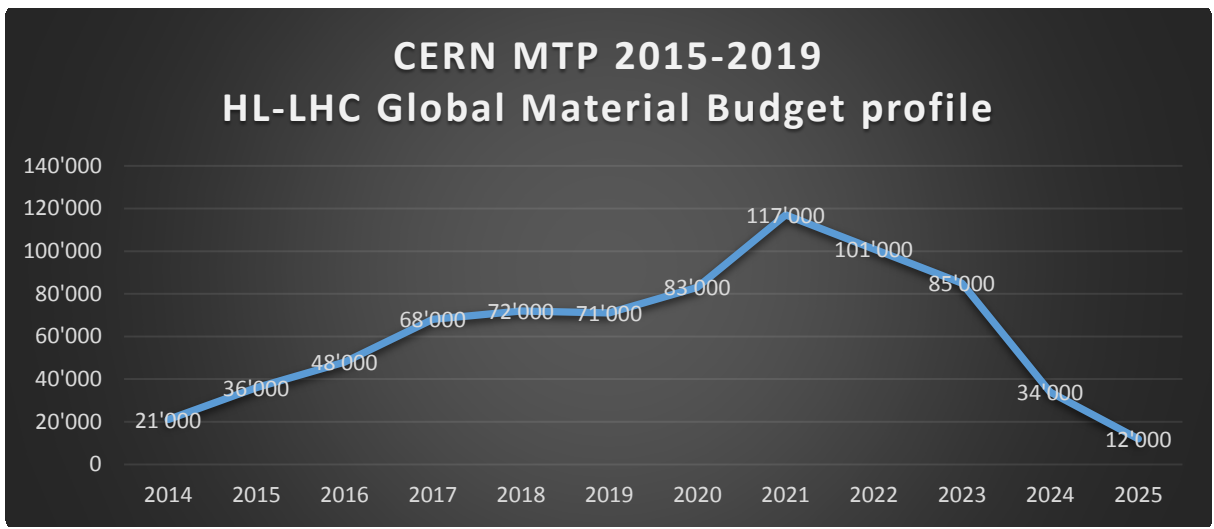

Figure 1-9: Budget allocation 2015 to 2025

Today the CERN draft budget attributes about 750 MCHF for the HL-LHC project until 2025, with certain assumptions of in-kind contributions from both the US and Japan. The discrepancy is not critical at this stage, since the modifications of certain systems are not yet fully defined. LHC operation at full energy and intensity will give important indications. The thorough investigation of potential synergy with the LHC consolidation project, together with various studies, should allow savings without compromising performance. Additional in-kind contributions to the hardware baseline would help alleviate the cost discrepancy and would also bring more personnel into the project.

A further possibility is to stage the project by using LS4, see Figure 1-7. Indeed the performance 'forecast' shown in Figure 1-7 is somewhat theoretical: there will be certainly a learning curve to pass from $2 \times 10^{34}$ cm$^{-2}$ s$^{-1}$ to (levelled) $5 \times 10^{34}$ cm$^{-2}$ s$^{-1}$, naturally favouring a staged approach. However, the 250 fb$^{-1}$ annual integrated luminosity goal can only be attained, and possibly even surpassed, when installation of all equipment is completed.

## 1.3 The collaboration

The LHC Luminosity Upgrade was conceived from the beginning as being even more international than the construction of the LHC machine, since US laboratories started to work on it with considerable resources well before CERN. In 2002–2003 collaboration between the US laboratories and CERN established the route for a machine upgrade [7]. The LARP programme was then setup and approved by the US Department of Energy (DOE). In the meantime, CERN was totally engaged in LHC construction and commissioning: it could only participate in Coordinated Accelerator Research in Europe (CARE), an EC-FP6 programme, in 2004–2008. CARE contained a modest programme for the LHC upgrade. Then two EC-FP7 programmes (SLH-PP and EuCARD) helped to reinforce the design and R&D work for the LHC upgrade in Europe, although still at a modest level. KEK in Japan, in the framework of the permanent CERN-KEK collaboration, from 2008 also engaged in activities for the LHC upgrade. LARP remained, until 2011, the main R&D activity in the world for the LHC upgrade.

Finally, with the approval of the EC-FP7 Design Study *HiLumi LHC* in 2011, and the maturing of the main project lines, the HL-LHC collaboration took its present form. It is worth noticing that FP7-HiLumi covers only the design of a few systems, given the limited amount of funding in such a programme. It has, however, allowed the formation and structuring of a European participation to the LHC Upgrade from the very beginning of the project. In 2014, CEA (Saclay, France), INFN (Milan and Genova, Italy) and CIEMAT (Madrid, Spain) have signed a further collaboration agreement to carry out design, engineering and prototyping work for HL-LHC magnets in addition to the FP7-EC commitment. In all three cases the CERN funding for the activities is approximately 50%, the rest being supported by the collaborating institutes. In Figure 1–10 a schematic indicating the various collaborating branches is shown.

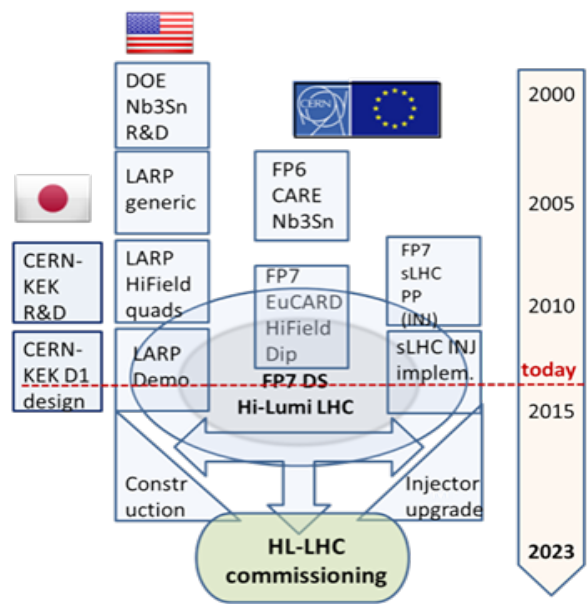

Figure 1-10: Timeline of the various collaboration branches, converging toward the LHC luminosity upgrade.

### 1.3.1 FP7-Hilumi LHC

The 'FP7 High Luminosity Large Hadron Collider Design Study' (FP7-HiLumi LHC) proposal was submitted in November 2010 to the EC Seventh Framework Programme. Approved with a full score of 15/15 it has been fully funded by the EC. The contract was signed by the fifteen partners (beneficiaries). KEK is a partner without EC funding – all of their funding is internal. The US laboratories were part of the proposal, without EC funding, but then for various reasons (mainly related to Intellectual Property issues) they could not sign the FP7-HiLumi LHC Consortium Agreement, thus they are external associates with no formal obligations. In practice LARP is excellently coordinated with FP7-HiLumi (see Section 1.3.2) and the project heavily relies on LARP to reach the project goals.

The workings of FP7 are such that each of the thirteen European institutions that are members of HiLumi LHC have to match the EC contribution with their own funding. In the case of FP7-HiLumi the matching funds equal the EC funds: each EU Institute receives 50% of the total cost (including overheads). The exception is CERN, which receives only 17% of its total costs, mainly for management and coordination. In Figure 1-11 the funding mechanism is explained. Given the success of the evaluation, see above, the project was ranked first in its category and was fully financed, with a EU contribution of M€4.9 against a request of M€4.97.

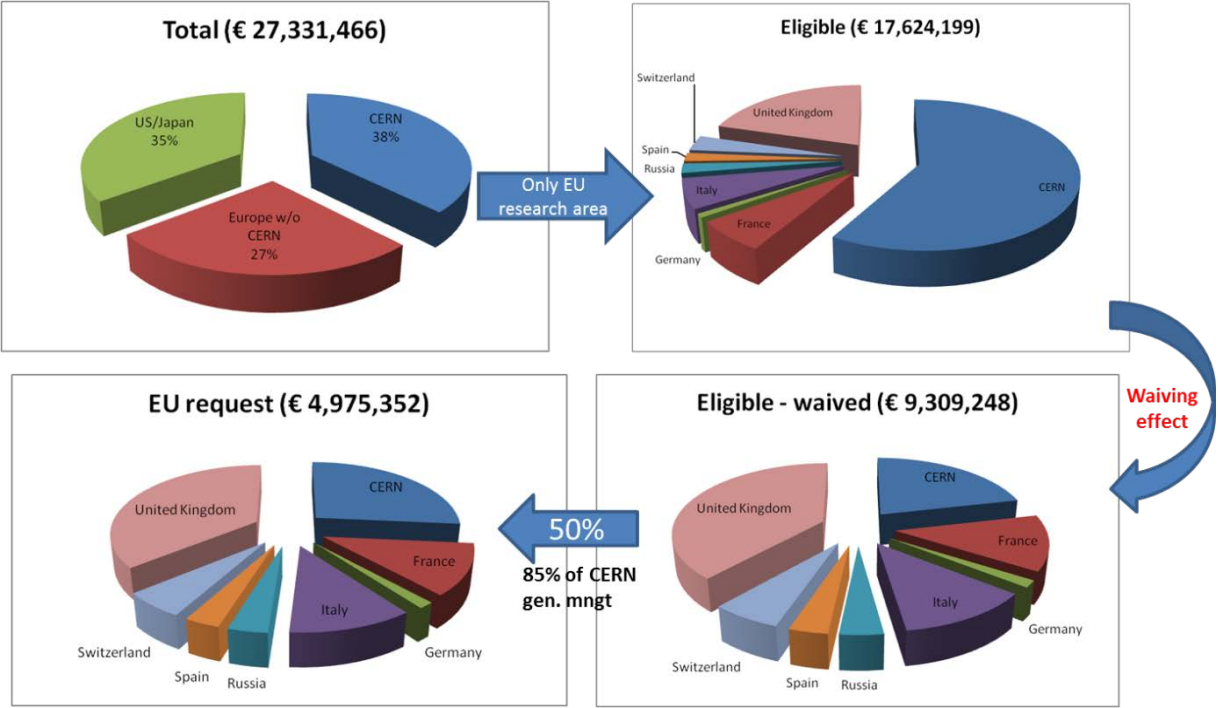

Figure 1-11: (a) Total estimation of the cost of the design study, subdivided by the US and Japan, EU institutes and CERN. (b) Total cost with the US and Japan removed (i.e. only costs that are eligible for funding by the EC). (c) Effect of CERN waiving the cost for technical works (recognizing that the HL-LHC is part of the core CERN programme financed via the normal budget), while keeping the extra cost generated by the management and coordination of the project. This is the total cost declared to the EC. (d) Cost claimed from the EC: 50% of the declared cost (eligible cost reduced by CERN waiving action).

In Figure 1-12 a list of the 15 FP7-HiLumi institutions is given, followed by a list of the five US collaborating institutes.

(a)

(b)

Figure 1-12: (a) Table showing the 15 members ('beneficiaries') of the FP7 HiLumi LHC design study and (b) the five LARP laboratories that are associated with the project.

1.3.2  LHC Accelerator R&D Program (LARP)

The LARP programme was initiated by the US Department of Energy (DOE) in 2003 to participate in the commissioning of the US-built interaction region triplets by bringing together and coordinating resources from the four US HEP laboratories (BNL, FNAL, LBNL and SLAC) with the inclusion of some universities as the programme evolved. By 2003 it was already recognized, based on the Tevatron experience, that an increase in LHC luminosity would become necessary after a decade of LHC operation to reduce the 'halving time', i.e. the time needed to reduce statistical errors by a factor of two. Consequently the programme focused – from the very beginning – on the design of improved focusing quadrupoles for the LHC low-$\beta$ insertion regions, finding a synergy with the various DOE high-field magnet (HFM) R&D programmes at the participating laboratories. The conductor of choice for this R&D programme was selected to be Nb$_3$Sn and therefore LARP became synergetic with another DOE programme, the Conductor Development Program (CDP), initiated in 1998 with the goal of improving the performance of Nb$_3$Sn. The LARP, CDP, and other US labs' HFM activities interacted in an extremely constructive way, achieving a substantial increase in the critical current performance of Nb$_3$Sn superconductors (Figure 1-13) and defining the assembly technique for accelerator quality high-field Nb$_3$Sn magnets in different kinds of configuration and with different apertures.

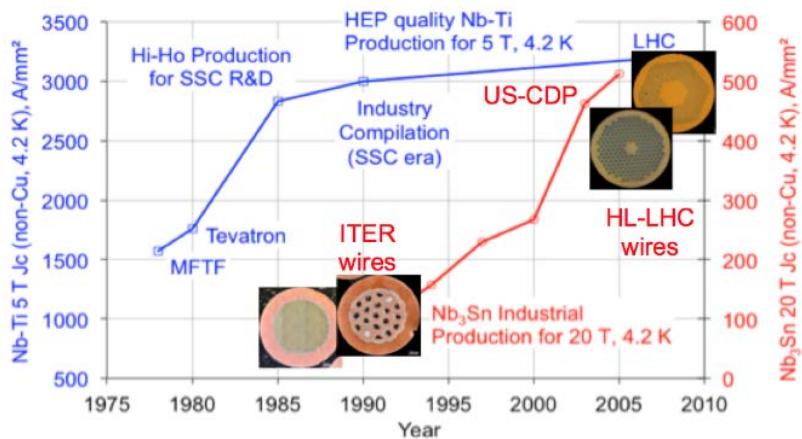

Figure 1-13: Left: Improvement in $J_c$ (current density) in $Nb_3Sn$ superconductor during the last three decades compared with Nb-Ti $J_c$ performance.

The LARP effort was funded at approximately $12–13 million/year, with 50% of the funding going directly to magnet development. Several magnets developed by LARP reached and surpassed the design field as shown in Figure 1-14(b) for one of the latest models (HQ02, a 120 mm aperture quadrupole assembled in 2014 and tested at FNAL and CERN). Additionally, LARP has demonstrated the scale-up of the $Nb_3Sn$ technology (i.e. the performance of the technology for magnets as long as 3 m) as shown in Figure 1-14(a) for the 90-mm aperture long quadrupole (LQ). The achievements of the US programmes, in particular LARP, but also of the general R&D high-field magnet programme, have led to the adoption of the $Nb_3Sn$ superconductor solution as the baseline for the HL-LHC's new focusing system and 11 T magnets.

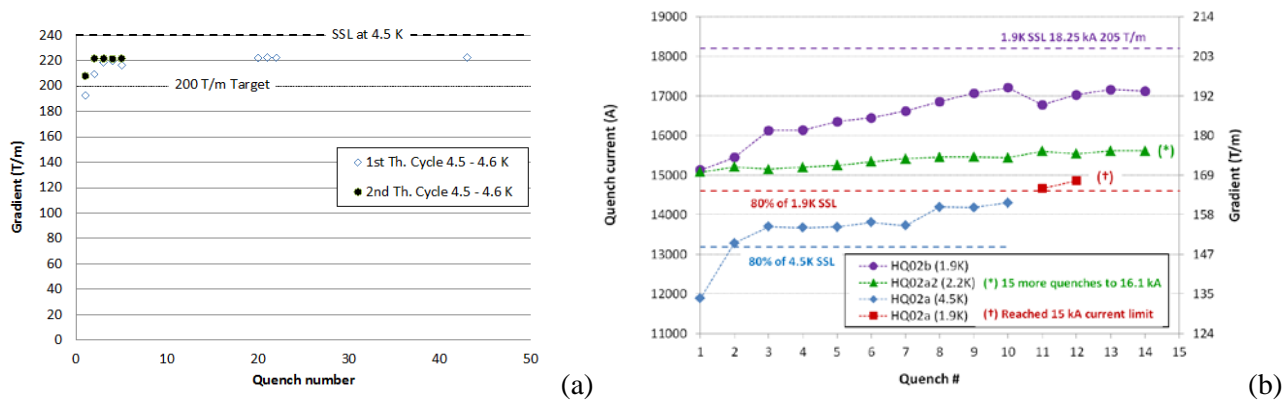

Figure 1-14: (a) Quench performance of the long quadrupole (LQ), the first quadrupole demonstrating the scale-up of $Nb_3Sn$ technology to lengths of interest for LHC applications (~3 m), (b) quench performance of HQ02 (120 mm aperture) after several re-assemblies, showing that in all cases the magnet achieved and passed the target (80% of the short sample limit (SSL)).

In addition to contributions to LHC commissioning and magnet development-related activities, LARP was also tasked with the support and promotion of accelerator physics R&D activities at the LHC accelerator. An additional very important aspect of this commitment was the institution of the Toohig Fellowship (http://www.interactions.org/toohig/) to support young accelerator physicists and engineers wishing to pursue research at the LHC in the early years of their career.

Recently, LARP has leveraged the superconducting RF capabilities and resources available at US laboratories and universities to focus on the development of crab cavities (Chapter 4), achieving transverse fields meeting the technical specifications for this system. In addition, a wide band feedback system is being researched and developed within LARP with the goal of mitigating transverse instabilities in the SPS and, possibly, in the LHC.

The DOE and CERN will negotiate the deliverables from the US in the coming years. In the CY15–CY17 period, LARP will concentrate on prototyping the elements needed by the HL-LHC project in which US national laboratories and universities have demonstrated excellent capabilities. In particular – subject to funding availability – LARP plans to build two short (1m) and three long (4 m) QXF magnet models to demonstrate the final design and reduce the risk during the construction period. In addition, LARP plans to deliver four fully-dressed SCRF crab cavities and a wide band feedback system prototype for tests in the SPS. This phase is expected to continue until the start of construction in the period 2018–2021.

### 1.3.3 KEK

Within the framework of the CERN-KEK collaboration, KEK has conducted $Nb_3Al$ superconductor R&D for the high-field magnets aimed at the future LHC upgrade from the early 2000s in collaboration with the National Institute of Materials Science (NIMS) in Japan. The $Nb_3Al$ superconductors are made by the rapid-heating, quenching transformation (RHQT) process, which was invented by NIMS. These superconductors have shown better critical current density and less strain dependence, and have been considered to be one of the promising candidates for high-field accelerator magnet applications. Nevertheless, KEK and NIMS faced technical difficulties in long wire production and it was judged in 2011 that the $Nb_3Al$ superconductor was unfortunately not ready for industrialization for the HL-LHC upgrade anticipated around 2022.

KEK has officially participated in the FP7 HiLumi LHC design study since 2011 in the context of enhancing the Japanese contribution to the physics outcome from the ATLAS experiment. Following the suppression of research activities on the development of the $Nb_3Al$ superconductor, the main effort was redirected to the conceptual design study for the beam separation dipole magnet, D1, situated immediately after the low-beta insertion quadrupoles in the HL-LHC machine. While the conceptual design study has been pursued dominantly by KEK, close collaboration with CERN and other partners has strengthened the success of the design study. The D1 magnet is based on the mature Nb-Ti technology. Design challenges are the tight control of the field quality with the large iron saturation, and the accommodation of the heat load and the radiation dose. The research engagement includes development of the 2 m long model magnet and testing at 1.9 K. KEK has also contributed to the HiLumi LHC design study through beam dynamics studies and the cooperative work associated with the crab cavity.

Aside from the FP7-HiLumi LHC, KEK has also participated in the LHC Injectors Upgrade (LIU) project. The main collaboration items have been consolidation and upgrade of PS Booster RF systems using Finemet-FT3L technology and development of the longitudinal damper system.

### 1.3.4 Other collaborations

In 2014, CEA (Saclay, France), INFN (Milan and Genova, Italy), and CIEMAT (Madrid, Spain), have each signed a further collaboration agreement to carry out design, engineering, and prototype work for HL-LHC magnets in addition to their FP7-HiLumi commitments. In all three cases, the CERN funding is about 50%, the rest being charged to the collaborating institutes.

#### 1.3.4.1 CEA

The CEA agreement concerns 'Research and Development for future LHC Superconducting Magnets'. It has six technical work packages, covering R&D for the HL-LHC and for post-LHC magnets. Among them, the following are of HL-LHC interest.

- Design and construction of a single aperture, 1 m long, full-size coil model magnet for the first quadrupole of the matching section, Q4. The magnet is based on classical Nb-Ti technology but has a very large aperture (90 mm) in a two-in-one cold mass, and thus presents a number of design challenges.

- Completion of the 13 T, large-aperture dipole Fresca2 (a technological HL-LHC work package that has served to promote $Nb_3Sn$ at CERN).

- Studies on Nb3Sn thermal properties and a finite element model of Nb3Sn cable.

### 1.3.4.2 INFN (Milan and Genova)

The INFN agreement is also related to R&D on superconducting magnets for the HL-LHC and concerns two main items:

- Design and construction of a prototype of each of the six high-order corrector magnets for the inner triplet, all with a single aperture of 150 mm. The work is based on Nb-Ti superferric technology and is carried out at INFN-LASA in Milano. An option based on the MgB2 superconductor is also being considered by INFN.

- Engineering Design of the superconducting recombination dipole magnet, D2, the first Two-in-One magnet, at the end of the common beam pipe. The work is based on Nb-Ti technology, with design challenges coming from the large aperture and the relatively high fields that have a parallel direction in both apertures. The work is being carried out at INFN-Genova.

### 1.3.4.3 CIEMAT (Madrid)

The CIEMAT agreement concerns the design and construction of a 1 m long prototype of the 150 mm aperture nested orbit corrector dipole for the inner triplet. It features two dipoles coils, rotated by 90° for simultaneous horizontal and vertical beam steering, in the same aperture. The main challenges are the mechanical structures to withstand the large torque, and the unusual force distribution arising when both field directions are needed.

## 1.4 Governance and project structure

Given the fact that the application for the FP7-HiLumi LHC Design Study marked the start of the project in its present form, the structure and terminology are borrowed from the typical FP7 style. To avoid any duplication the governance of the whole HL-LHC project is conceived as an extension of the governance that has been instituted for the governance of the FP7-HiLumi LHC.

As noted above, the FP7-HiLumi LHC covers only a few work packages (WPs), although they are the backbone of the upgrade. The WP structure, with tasks arranged in a tree-like structure, is the basic arrangement of the project. LARP is a parallel structure, independently funded, associated with FP7 with connections both at project management level as well as at WP/task level to maximize synergy. KEK is a direct member of FP7-HiLumi. It is worth noting that HiLumi LHC is the term indicating the part of the HL-LHC that is covered by FP7 funds, even if in practice it has become a popular name for the whole project. In Figure 1-15 the general governance of the project is shown. Each body contains the FP7 part and the part that is not covered by FP7. The Steering Committee is the main managing body: it meets regularly every two months and all WPs are represented there, with the addition of the LARP representatives. It oversees the progress of the technical work and planning, approving the milestones and deliverables. The Steering Committee usually meets in its 'enlarged' form, including the WPs not covered by FP7 and including the LARP leadership. The Collaboration Board is the highest-level governance body with representation from each institute.

In the case of approval of formal FP7 acts, only the FP7-WP coordinators and FP7 Institutes can vote. It is worth noting that the collaboration is based on a consortium agreement, signed by the 15 members (in FP7 terminology, beneficiaries) of FP7-HiLumi LHC. The US laboratories are not members of FP7-HiLumi LHC, however representatives of each US laboratory, including the LARP director, are co-opted onto the enlarged Collaboration Board. The formal link with the US laboratories is assured by the recently signed CERN-DOE Protocol II concerning the LHC and its upgrades. Given the fact that CERN is responsible for the LHC machine, the CERN director general, through his representative in the Collaboration Board, the project coordinator, has the right of veto.

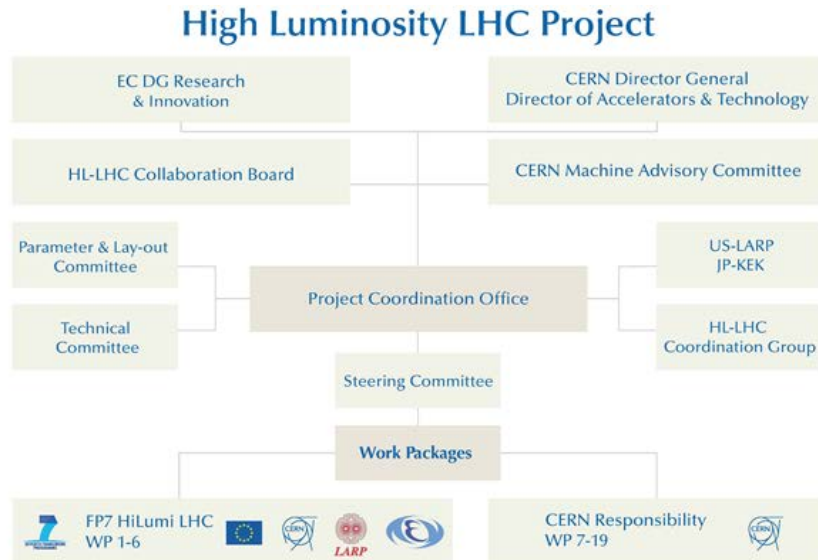

Figure 1-15: The general governance scheme of FP-7 HiLumi LHC, used for the whole HL-LHC project (see text for details)

The Parameter and Layout Committee and the Technical Committee have mainly technical functions inside the project. The Coordination group, chaired by the HL-LHC leader, constitutes the meeting point between CERN Management, HL-LHC, LIU and Detector Management.

A new structure, more suited to a project that is passing from the design study phase to construction project status, is under study and will be operative from November 2015 when the FP7 consortium agreement comes to an end.

In Figure 1-16 the project structure, with all WPs and their coordinators, as well as the main collaborators, is shown. Typically, each WP is assigned three to six tasks. The tasks are the core of the technical work.

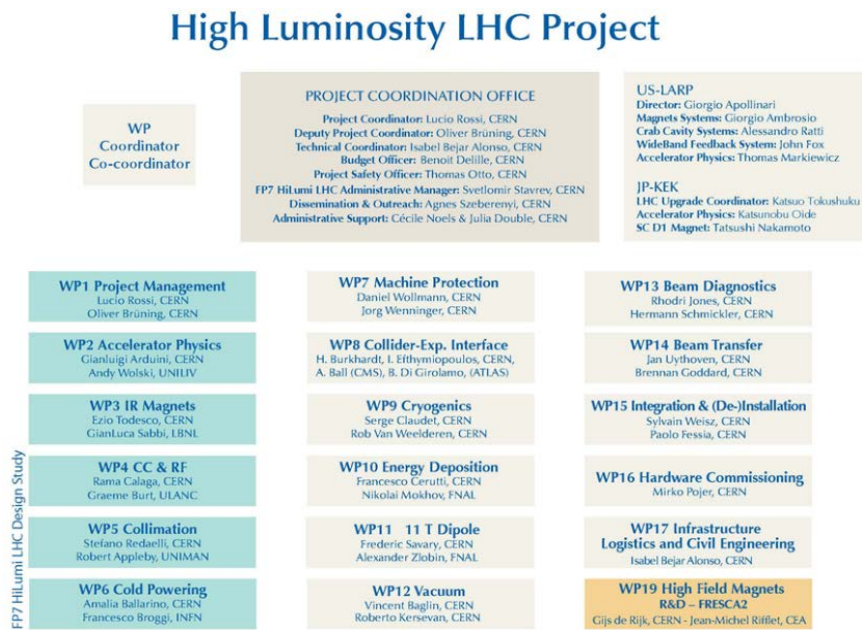

Figure 1-16: HL-LHC project structure, with FP7 part indicated in dark green. The orange box refers to the high-field magnets work package, which was started before the HL-LHC in the framework of generic R&D for the LHC upgrade.